\documentclass[conference]{IEEEtran}
\IEEEoverridecommandlockouts

\usepackage[nolist,nohyperlinks]{acronym}
\def\BibTeX{{\rm B\kern-.05em{\sc i\kern-.025em b}\kern-.08em
    T\kern-.1667em\lower.7ex\hbox{E}\kern-.125emX}}

\usepackage{graphics}
\usepackage{graphicx}
\usepackage{xcolor}
\usepackage{amssymb}
\usepackage{multirow}
\usepackage{url}
\usepackage{soul}
\usepackage{etoolbox, xspace}
\usepackage[figurename=Figure]{caption}
\usepackage{subfigure}
\usepackage{framed}
\usepackage{cite}
\usepackage{algorithm} 
\usepackage{algpseudocode} 
\usepackage[nolist,nohyperlinks]{acronym}
\usepackage{hyperref}
\usepackage[left=0.68in,right=0.68in,bottom=1in, top=0.72in]{geometry}
\begin{document}
\def\thetitle{Cross-Domain AI for Early Attack Detection and Defense Against Malicious Flows in O-RAN}
\title{\thetitle}

\author{
\IEEEauthorblockN{
    Bruno Missi Xavier \IEEEauthorrefmark{2}\IEEEauthorrefmark{3},
    Merim Dzaferagic\IEEEauthorrefmark{3},
    Irene Vilà\IEEEauthorrefmark{3}\IEEEauthorrefmark{4},
    Magnos Martinello\IEEEauthorrefmark{2}, and
    Marco Ruffini\IEEEauthorrefmark{3}
}

\IEEEauthorblockA{\IEEEauthorrefmark{2}
    Federal University of Espírito Santo, Espírito Santo, Brazil}
\IEEEauthorblockA{\IEEEauthorrefmark{3}
    Trinity College Dublin, Ireland}
\IEEEauthorblockA{\IEEEauthorrefmark{4}
    \if 0 Signal Theory and Communications Department of \fi Universitat Politècnica de Catalunya, Barcelona, Spain} 
}
\maketitle

\newcommand{\SDNClass}{\textit{In-Network Classifier}\xspace}
\newcommand{\SDRClass}{\textit{RAN Classifier}\xspace}
\newcommand{\benchmark}{\textit{Benchmark Classifier}\xspace}
\newcommand{\SDNDataset}{\textit{In-Network Dataset}\xspace}
\newcommand{\SDRDataset}{\textit{\acs{ran} Dataset}\xspace}

\newcommand{\NAME}{\textit{{\color{red}  NAME}}\xspace}

\begin{acronym}
  \acro{ns}[NS]{Network Softwarization}
  \acro{sdn}[SDN]{Software-Defined Network}
  \acro{sdr}[SDR]{Software-Defined Radio}
  \acro{o-ran}[O-RAN]{Open Radio Access Network}
  \acro{kpi}[KPI]{Key Performance Indicator}
  \acro{pm}[PM]{Performance Measurement}
  \acro{bs}[BS]{Base Station}
  \acro{ue}[UE]{User Equipment}
  \acro{rt}[RT]{Real Time}
  \acro{ric}[RIC]{RAN Intelligent Controller}
  \acro{mac}[MAC]{Medium Access Control}
  \acro{phy}[PHY]{Physical}
  \acro{near-rt}[near-RT]{Near-Real Time}
  \acro{ru}[RU]{Radio Unit}
  \acro{du}[DU]{Distributed Unit}
  \acro{cu}[CU]{Centralized Unit}

  \acro{ml}[ML]{Machine Learning}
  \acro{ai}[AI]{Artificial Intelligence}
  \acro{k-nn}[$k$-NN]{$k$-Nearest Neighbors}
  \acro{ann}[ANN]{Artificial Neural Network}
  \acro{svm}[SVM]{Support Vector Machine}

  \acro{iot}[IoT]{Internet of Things}
  \acro{ddos}[DDoS]{Distributed Denial-of-Service}
  \acro{voip}[VoIP]{Voice over IP}
  \acro{usrp}[USRP]{Universal Software Radio Peripheral}
  \acro{vm}[VM]{Virtual Machine}
  \acro{ids}[IDS]{Intrusion Detection System}
  \acro{ran}[RAN]{Radio Access Network}

\end{acronym}
\begin{abstract}
In the fight against cyber attacks, \ac*{ns} is a flexible and adaptable shield, using advanced software to spot malicious activity in regular network traffic.
However, the availability of comprehensive datasets for mobile networks, which are fundamental for the development of \ac{ml} solutions for attack detection near their source, is still limited. Cross-Domain \ac*{ai} can be the key to address this, although its application in \ac*{o-ran} is still at its infancy.
To address these challenges, we deployed an end-to-end \acs*{o-ran} network, that was used to collect data from the \acs*{ran} and the transport network. These datasets allow us to combine the knowledge from an in-network \ac*{ml} traffic classifier for attack detection to bolster the training of an \acs*{ml}-based traffic classifier specifically tailored for the \acs*{ran}. Our results demonstrate the potential of the proposed approach, achieving an accuracy rate of 93\%. This approach not only bridges critical gaps in mobile network security but also showcases the potential of cross-domain \acs*{ai} in enhancing the efficacy of network security measures.
\end{abstract}

\vspace{-0.3cm}
\section{Introduction}\label{sec:intro}
Cyber attacks are on the rise \cite{chinese2023world}, and networks are on the front lines of defense. Switches, routers, servers, and end-users all need protection from malicious threats. \ac{ns} has become a critical tool in this fight, offering flexibility, scalability, and the ability to quickly deploy cutting-edge software solutions. \ac{ns} helps security professionals to identify malicious activity in a sea of benign network traffic. The ability to adapt and respond quickly to new threats is essential in the fight against cyber adversaries. Therefore, \ac{ns} enables resilience and integrity of modern network infrastructures \cite{popescu2022network}.

In terms of \ac{ns}, \ac{sdn} has brought in a new era of advanced programmability. Among other features, it allows for the integration of \ac{ml} into the data plane \cite{kaur2021review, qin2020learning,qin2022bringing}. Programmable network devices have revolutionized various aspects of networking, enabling \ac{ml}-based dynamic congestion control strategies \cite{sha2023machine, sacco2023hint}, intelligent load balancing mechanisms \cite{liu2023load,alhilali2023artificial}, and precise Quality of Service (QoS) management \cite{wu2023p4sqa,keshari2021systematic,ali2022qos}. A number of recent publications studies traffic classification \cite{zhu2022machine, paramasivam2023cor, xavier2022map4, musumeci2022machine,zang2023towards}, where the works in \cite{xavier2022map4, musumeci2022machine,zang2023towards} use traffic classification for attack detection. Even though programmable data planes are widely used, there are still challenges to consider when developing and deploying new features. While the P4 language offers immense potential, limitations such as the lack of support for floating-point operations and restricted loop controls must be acknowledged \cite{sacco2023p4fl}. Moreover, programmable devices face constraints related to the available memory and processor cycles needed for seamless packet forwarding at line rate. These limitations restrict complex \ac{ml} models like \ac{svm} and \ac{ann}, emphasizing the need for innovative solutions to harness programmable network capabilities fully\footnote{\url{https://arxiv.org/pdf/2308.00797.pdf}}.

\begin{figure*}[!t] 
\centering
\includegraphics[scale=0.6]{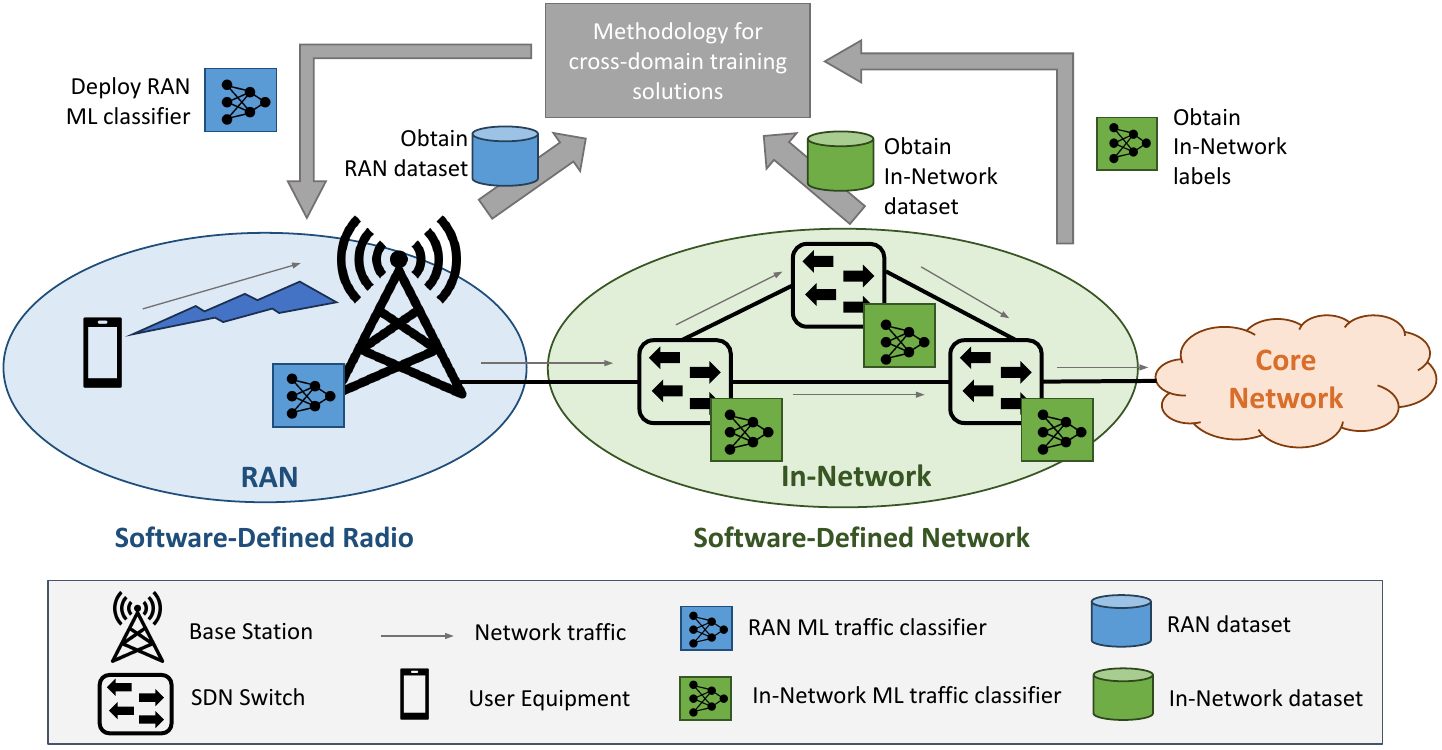}
\vspace{-0.1cm}
\caption{System overview.}
\label{fig:overview}
\vspace{-0.6cm}
\end{figure*}

Due to attacks targeting networking equipment, cellular networks are facing new challenges in ensuring continuous service availability. To address these challenges, upcoming fifth (5G) and sixth (6G) generations of cellular networks are undergoing a fundamental shift from rigid and monolithic structures to agile and disaggregated architectures, driven by softwarization \cite{bonati2021intelligence}. This transformation enables the development of innovative methods for the early detection and precise localization of attack sources, enabling network operators to promptly intercept and halt malicious flows before they can permeate the network. 
A wide range of literature is available on diverse security datasets for the transport network. These datasets are carefully prepared and include packets alongside their corresponding class labels \cite{ferrag2022edge,vamsi2022detailed}. On the other hand, traffic classification and attack mitigation in \ac{o-ran} is still a relatively new area of research. Recent research has explored how the \ac{phy} and \ac{mac} layers can contribute to the creation of advanced \ac{ml} models \cite{xavier2023ICC} to identify attacks early. However, a notable gap persists in the form of comprehensive datasets encompassing representative features for mobile networks, highlighting the critical need for further research and data collection efforts in this evolving landscape. To bridge this gap, researchers have turned to using datasets from other domains, such as transport network\cite{pinheiro20235geco}, and core network\cite{10195597}, to support their work on mobile networks. The logical next step is to extend these endeavors to a cross-domain \ac{ai} approach. Cross-domain \ac{ai} in \ac{o-ran} is a novel approach to network management and optimization that utilizes \ac{ai} to integrate data and insights from different domains, like the \ac{ran} and the transport network \cite{alliance2021ran_WG2}. This enables network operators to gain a more comprehensive view of their networks and to make more informed decisions about how to allocate resources and mitigate threats.


To fill the abovementioned research gap, the research in this paper is based on realistic data collected in our OpenIreland testbed facilities\footnote{\url{http://openireland.eu}}. Considering that the in-network traffic classification (i.e. transport network \ac{sdn} classification) is well-studied, the feedback from the transport network has the potential to improve the traffic classification in the \ac{ran}. Therefore, the focus of our research revolves around the concept of cross-domain \ac{ai} in \ac{o-ran}. We train and deploy \ac{ml} classifiers as \emph{xApps} running on the \ac{near-rt} \ac{ric}. These classifiers are trained on data collected from the \ac{ran}, and feedback received from the transport network. This cross-domain \ac{ai} approach represents a significant departure from conventional approaches that rely on in-network traffic classification and attack detection. Besides the clear benefit related to attacks being detected and stopped at the source of the attack, it enables model updates and continuous improvement of \ac{ran} classifiers based on feedback received from the transport network. The contributions of this work can be summarized as follows:
\begin{itemize}
    
    \item We conduct experiments to study the impact that a cross-domain \ac{ai} approach has on enhancing security measures within a realistic end-to-end \ac{o-ran} deployment;
    \item We present and demonstrate a cross-domain \ac{ai} methodology for training a \ac{ran} \ac{ml} classifier based on automatic labeling of \ac{kpi} \ac{pm} datasets leveraging in-network traffic classification;
    \item We evaluate the ongoing performance improvement of traffic classifiers deployed as \emph{xApps} on a \ac{near-rt} \ac{ric}, which demonstrate the continuous training of new \ac{ml} models based on feedback received from in-network traffic classifiers.
\end{itemize}

We emphasize that, given the nature of mobile networks, early attack detection plays a pivotal role in safeguarding adjacent infrastructure against potential flood from malicious flows. Such preemptive measures are crucial, as they can prevent interruptions in critical services within the network infrastructure. 
\begin{figure*}[!t] 
\centering
\includegraphics[scale=0.6]{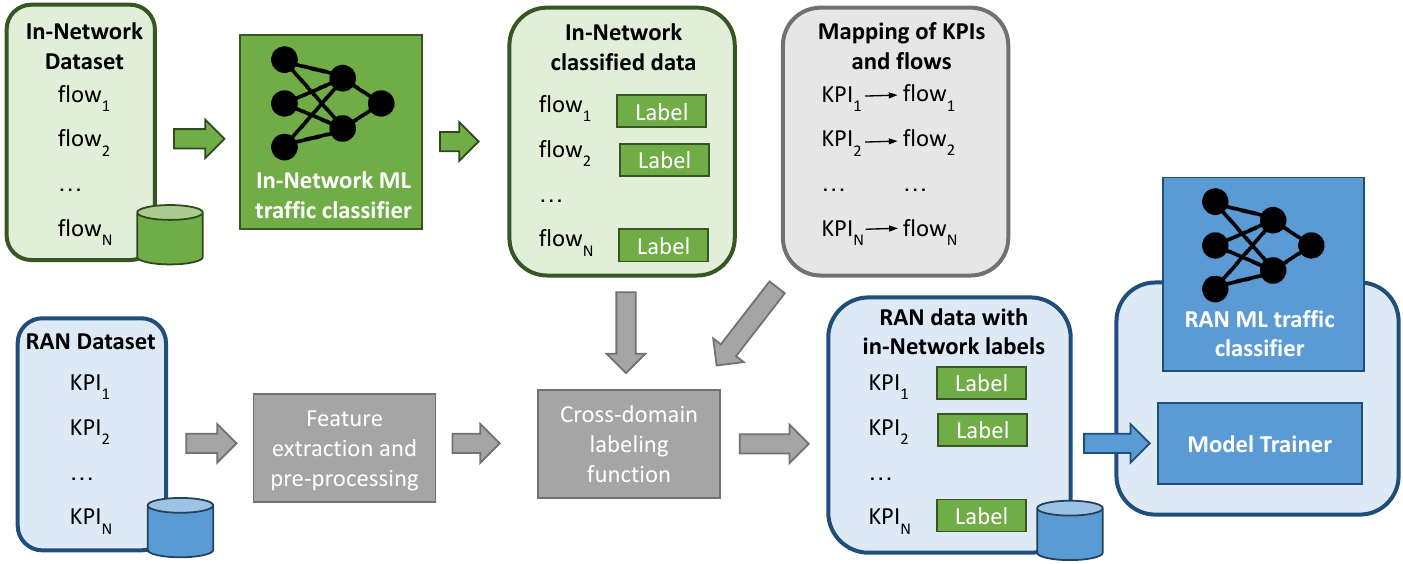}
\vspace{-0.1cm}
\caption{Cross-domain methodology scheme.}
\label{fig:proc}
\vspace{-0.6cm}
\end{figure*}
\vspace{-0.1cm}
\section{Cross-Domain AI for \ac{ml} Feedback and Training Process}\label{sec:cross-domain}
\vspace{-0.1cm}
This section provides a description of the cross-domain \ac{ai} methodology for improving the detection and various types of attacks in the \ac{ran}. We collected information from two domains, namely the \ac{ran} and the transport network. Considering that the transport network has more information at its disposal for traffic classification (i.e., packet headers and flows), we will rely on these classifiers to continuously refine the traffic classifiers in the \ac{ran}. 

Figure~\ref{fig:overview} depicts the overall system. There are three distinct parts of the system: (1) \ac{ran}, (2) in-network, and (3) cross-domain \ac{ai} methodology. The \ac{ran} is composed of an \ac{o-ran} deployment which includes a 5G \ac{bs}, i.e., \ac{ru}, \ac{du}, \ac{cu} and a \ac{near-rt} \ac{ric}, that incorporates a \ac{ran} \ac{ml} traffic classifier for attacks detection. The in-network refers to the transport network which consists of switches with in-network \ac{ml} traffic classification. The cross-domain \ac{ai} methodology refers to our approach to consolidate the information from the first two system parts to improve the overall attack detection accuracy.



In terms of traffic classification, a fundamental \ac{ml} challenge is the identification of the features that describe the classes accurately. In \ac{sdn}, traffic analysis traditionally focuses on two levels of granularity: individual packets and aggregated flows. The features are derived either intrinsically from data, such as headers and payloads, or extrinsically, involving factors such as packet arrival time. The selection of the appropriate granularity and features is pivotal and varies significantly depending on the specific application at hand. This decision-making process is widely explored in \cite{xavier2022map4, nguyen2020efficient,zhou2020flow, barradasflowlens}. 

On the other hand, the \SDRClass has no access to the features related to data flows and packet headers. It has to rely on \ac{ran} \acp{kpi} to extract information about the traffic classes. The range of \acp{kpi} includes but is not limited to \ac{ue} measurements, E2 Node Measurements, and E2 Node Load-related metrics, as clarified in \cite{alliance2021ran_WG3_RC}. These \acp{kpi} are commonly used to estimate channel quality and to allocate resources to \acp{ue} connected to the \ac{bs}. In \cite{xavier2023ICC} the authors prove that these features do carry some information about the traffic, and therefore can be used for traffic classification. However, compared to the \ac{sdn} in-network classification, the \ac{ran} classifiers are limited in their understanding of different traffic classes. Therefore, these two domains are an excellent choice to demonstrate the potential of cross-domain \ac{ai} in an \ac{o-ran} deployment.


Figure~\ref{fig:proc} depicts the proposed methodology that allows us to consolidate the measurements and predictions in two different network domains, i.e., the \ac{ran} and transport network. It is important to notice that the primary challenge lies in establishing a robust connection between the features used for classification in both domains. This challenge arises due to time granularity differences in the feature collection in different network domains, i.e., traffic flows and \ac{ran} \acp{kpi}. Our approach links the two domains through the cross-domain labeling function (see Figure~\ref{fig:proc}). First, this function identifies the \emph{\ac{ue} IP} and defines a measurement time window in the \ac{ran} part of the network. Then this time window is associated with traffic flows in the transport network, and filtered for the \ac{ue} of interest. 

As shown in Figure~\ref{fig:proc}, when new traffic arrives, the near-\ac{rt} \ac{ric} within the \ac{o-ran} framework will perform the \emph{Feature extracting and preprocessing}. That results in an unlabeled \ac{kpi} dataset, i.e., \SDRDataset. From the \ac{ran}, the traffic flows through the transport network, where it undergoes classification by the \SDNClass. The inferred labels are combined with the features used for flow classification in the transport network and stored as part of the \SDNDataset. Then, the \emph{Cross-Domain labeling function} comes into play by connecting one or more instances of the \SDRDataset with one or more \SDNDataset samples. Finally, this integration results in the labeling of the previously unlabeled \SDRDataset.

The next step in the process involves training a new \SDRClass model intended for deployment on the near-\ac{rt} \ac{ric} of the 5G \ac{bs} as an \emph{xApp}. Once a model is deployed, different triggers can be used to initiate new model training and deployment in the \ac{ran}, e.g., number of new samples stored in the \SDRDataset, change in traffic distribution over time, time since last training, model accuracy depreciation.

The training of the \ac{ran} classifier is powered by \mbox{AutoML}\footnote{\url{https://www.automl.org/automl/}}. As the name suggests, \mbox{AutoML} is a tool designed to automate \ac{ml} processes. This automation significantly enhances the efficiency of \ac{ml} procedures by automating key steps such as selecting and constructing appropriate features, identifying suitable model families, optimizing model hyperparameters, and post-processing \ac{ml} models. By leveraging AutoML, we transform the model training and validation process into a zero-touch operation, reducing the necessity for human intervention.

Upon deployment of the \SDRClass, the new model engages in identifying traffic classes at the very edge of the network, i.e., at the \ac{bs}. This allows us to identify and stop malicious traffic near its source. The process of collecting samples in both domains continues after the deployment of models, which allows us to continuously improve the models accuracy.
\section{Case Study: Intrusion Detection}\label{sec:result}

\newcommand{\bgcell}[2][c]{%
  \textbf{\begin{tabular}[#1]{@{}c@{}}#2\end{tabular}}}
\begin{table}[]
    \centering
    \caption{Datasets  Summary}
    \label{tab:dtst}
    \vspace{-0.1cm}
    \begin{tabular}{|l|c|c|} \hline 
         \bgcell{Class}&  \bgcell{In-Network Training \\Dataset}& \bgcell{RAN Training\\Dataset}\\ \hline 
          Web Browsing& 50,068 & 139,646\\ 
         VoIP& 8,200 & 42,773\\ 
         IoT& 90,040 & 21,856\\ 
         YouTube& 13,230 & 13,337\\ 
         DDoS-Ripper& 111,735 & 11,974\\ 
         DDoS-Hulk& 125,302 & 17,401\\ 
         PortScan& 31,594 & 24,830\\  
         Slowloris& 82,399 & 57,142\\ \hline
    \end{tabular}

\end{table}

We start this section by describing the datasets employed for the experiments (Section~\ref{sec:case_study:dataset}) and outlining the specifics of our experimental setup in Section~\ref{sec:case_study:setup}. In Section~\ref{sec:case_study:lifecycle}, we provide insights into how the accuracy of the classifier evolves over time. Finally, we also showcase the performance of our Cross-Domain \ac{ai} methodology (Section~\ref{sec:case_study:class_result}. 
\subsection{Dataset forming}\label{sec:case_study:dataset}
In the construction of our \SDNDataset, we deliberately chose to operate at the flow granularity level, due to the outstanding model accuracy. This performance has been documented by the authors of \cite{xavier2022map4, barradasflowlens}. The dataset is created by aggregating packets based on a specific flow key. This key includes features such as source and destination IP addresses, source and destination ports, as well as the IPv4 Protocol extracted from the packet's header. The list of features is outlined in \cite{xavier2022map4}.

In contrast, the \SDRDataset harnesses the \acp{pm} available to the near-\acp{rt} \acp{ric}. \cite{xavier2023ICC} offers a comprehensive insight into this dataset’s structure and outlines the feature selection process. The authors also discuss the challenges related to traffic classification with features from the \ac{phy} and \ac{mac} layers. To support the cross-domain \ac{ai} methodology, we introduce the \ac{ue} IP address and the arrival time of each collected sample as extrinsic features in both datasets, i.e., \SDRDataset and \SDNDataset. These datasets were collected as part of our experiments, leveraging the resources of our OpenIreland testbed.

\begin{figure}[] 
\centering
\includegraphics[scale=0.5]{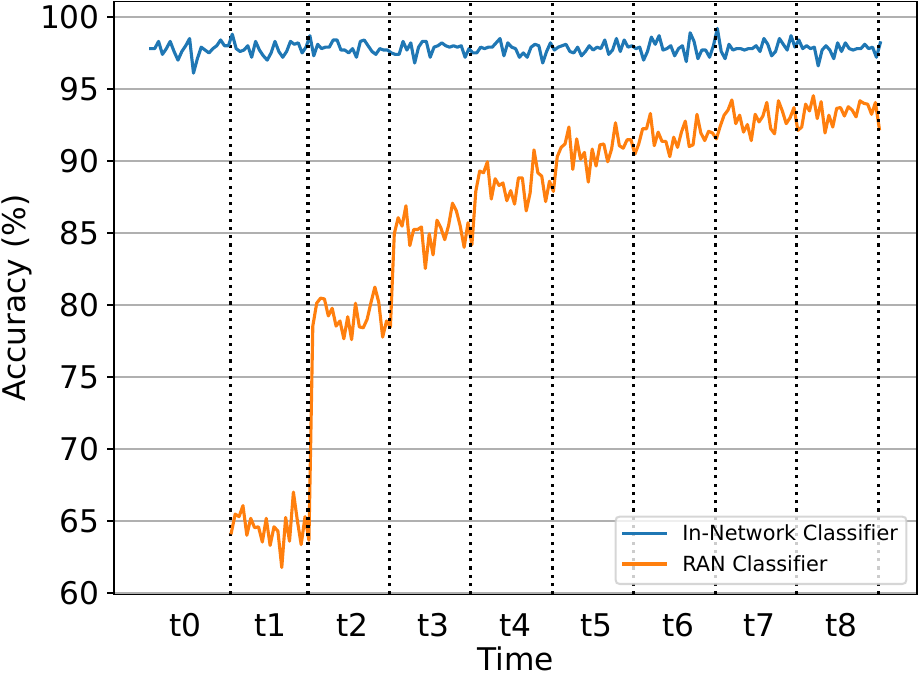}
\vspace{-0.1cm}
\caption{Temporal accuracy trends for continuous Cross-Domain \ac{ai} operation.}
\label{fig:lifecycle}
\vspace{-0.3cm}
\end{figure}


The traffic pattern generation relies on a set of established Traffic Generator Tools detailed in \cite{xavier2023ICC, sharafaldin2018toward}. The generated traffic classes can be divided into two primary types: \emph{Benign} and \emph{Attack}. Each category consists of the following traffic types:

\begin{itemize}
    \item \emph{Benign}: \emph{Web Browsing}, \emph{\ac{voip}}, \emph{\ac{iot}}, and \emph{YouTube};
    \item \emph{Attack}: \emph{\ac{ddos} Ripper}, \emph{\ac{ddos} Hulk}, \emph{PortScan}, and \emph{Slowloris}. 
\end{itemize}

\begin{table*}
    \centering
    \caption{Summary of classification results}
    \label{tab:class_result}
    \vspace{-0.1cm}
    \begin{tabular}{rccc|ccc|ccc} \hline\hline  
        &  \multicolumn{3}{c|}{\benchmark}  &  \multicolumn{3}{c|}{\SDNClass}&  \multicolumn{3}{|c}{\SDRClass trained for t8}\\  \cline{2-10}  
 & Prediction& Recall& F1-Score & Prediction& Recall& F1-Score& Prediction& Recall&F1-Score\\  \cline{2-10} 
 Web Browsing& 0.98& 0.99&0.98& 0.96& 0.97& 0.96 & 0.96& 0.97& 0.96\\  
 VoIP& 0.97& 0.98&0.97& 1.00& 1.00& 1.00 & 0.95& 0.96& 0.96\\ 
 IoT& 0.97& 0.96&0.97& 1.00& 1.00& 1.00 & 0.96& 0.95& 0.95\\  
 YouTube& 0.91& 0.88&0.89& 0.98& 0.80& 0.88 & 0.78& 0.74& 0.76\\ \hline  
 DDoS-Ripper& 0.90& 0.88&0.89& 0.97& 0.96& 0.97 & 0.83& 0.82& 0.82\\  
 DDoS-Hulk& 0.91& 0.90&0.91& 0.99& 1.00& 0.99 & 0.86& 85& 0.86\\   
 PortScan& 0.95& 0.93&0.94& 0.99& 1.00& 1.00 & 0.92& 0.91& 0.91\\   
 Slowloris& 0.94& 0.95&0.95& 0.98& 0.99& 0.98 & 0.92& 0.92& 0.92\\ \hline \hline

    \end{tabular}
\vspace{-0.4cm}

\end{table*}

\begin{figure*}[] 
\centering
\subfigure[\benchmark]{\includegraphics[scale=0.45]{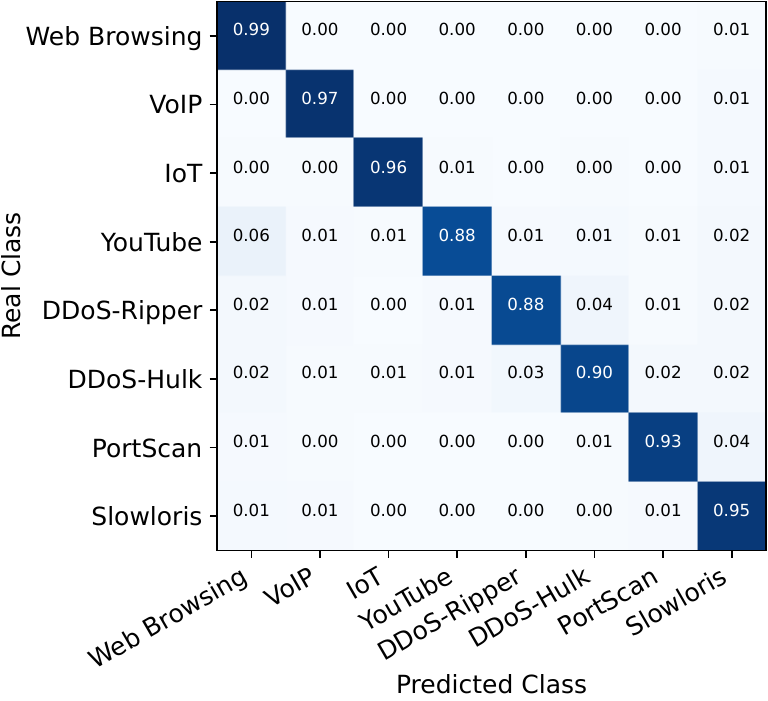}\label{fig:artificial_cm}}
\subfigure[\SDNClass.]{\includegraphics[scale=0.45]{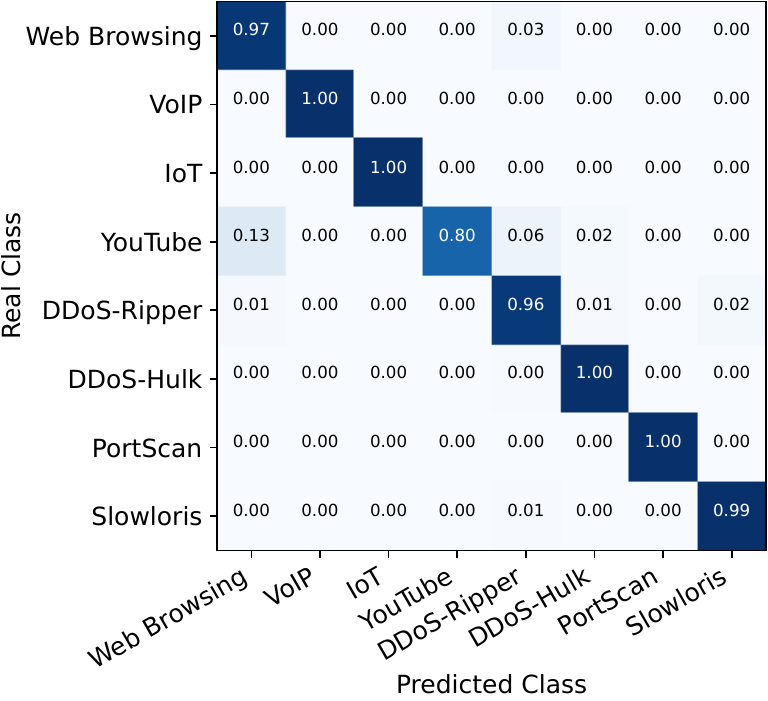}\label{fig:sdn_cm}}
\subfigure[\SDRClass.]{\includegraphics[scale=0.45]{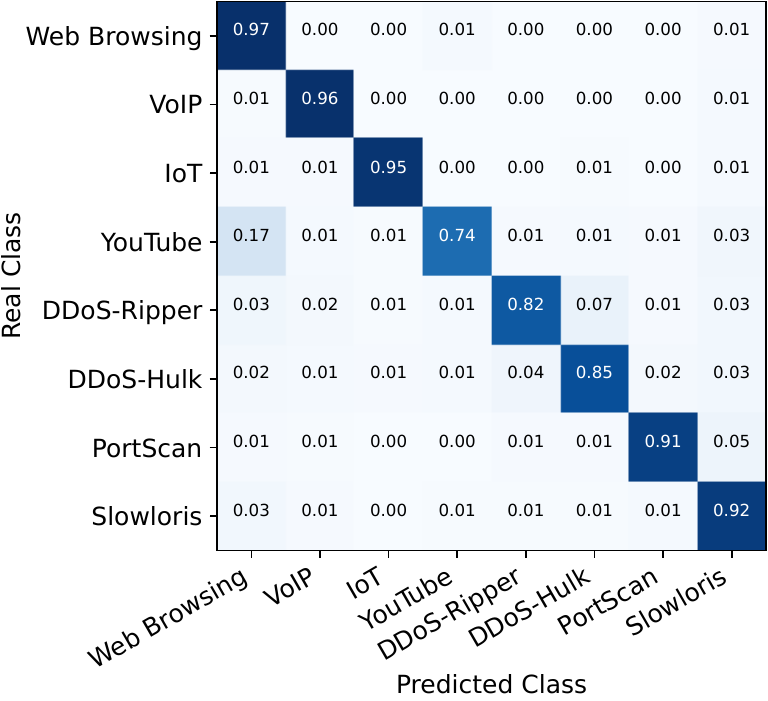}\label{fig:sdr_cm}}
\vspace{-0.1cm}
\caption{Confusion Matrix of Traffic Classification.}
\label{fig:classConfMatrix}
\vspace{-1.6em}
\end{figure*}

In Table~\ref{tab:dtst}, we provide an overview of our datasets derived through two distinct phases. The column labeled \emph{In-Network Training Dataset} comprises data specifically gathered for training the \SDNClass. Instead, the column labeled \emph{RAN Training Dataset} includes the data for training the \SDRClass. 


\subsection{Experimental Setup} \label{sec:case_study:setup}
As shown in Figure~\ref{fig:overview}, our experimental setup consists of an \ac{o-ran} deployment with a transport and core network. As highlighted earlier, we combine known approaches for traffic classification in the transport and \ac{ran} network to study the potential of a cross-domain approach that will improve the overall system accuracy. 

\subsubsection{\textbf{\ac{ran} deployment}}
Our \ac{ran} deployment is based on the work published in \cite{xavier2023ICC}. We rely on Ettus Research \acp{usrp} running the \emph{srsRAN} virtualized \acp{bs}. Our experimental setup is composed of five \acp{usrp}, with four of them working as \acp{ue} and one acting as a \ac{bs}. For the purpose of our experiments, we have also developed a near-\ac{rt} \ac{ric} that enables closed-loop control through \emph{xApp} deployments. These applications enable the implementation of diverse control strategies, leveraging inputs from measurements to make decisions based on optimization algorithms, \ac{ml} models, or predefined policies. 

\subsubsection{\textbf{In-network deployment}}
Our in-network scenario is based on the specifications outlined in \cite{xavier2022map4}. The approach involves converting a Decision Tree model into a straightforward \emph{if-else} chain. 
We use a single server to host our \SDNClass, positioned between the \ac{ran} and the broader network infrastructure. This allows us to route all network traffic through the switch hosting the \SDNClass.
\subsection{Cross-Domain AI Continuous Training and Operation} \label{sec:case_study:lifecycle}

To assess the accuracy of the proposed Cross-Domain \ac{ai} methodology over time, the \SDNClass is initially trained by using the whole \emph{In-Network Training Dataset} in Table~\ref{tab:dtst}. In our experimental setup, this allows us to make the earlier mentioned assumption that the \SDNClass is aware of all types of attacks. The reader is reminded that such classifiers are commonly updated as part of network security updates. Our goal is now to test whether our cross-domain approach will allow us to improve the \SDRClass over time. Therefore, we adopted an iterative process. Once the \SDNClass is trained, we deploy it in $t0$. This classifier remains active throughout the entire online evaluation period and remains frozen (i.e. it is not re-trained). 



Please note that during the initial phase, denoted as $t0$, the \SDRClass does not exist. This marks the outset of our data collection process. Upon labeling a predetermined number of instances (in this case, we used $36,500$ instances as a threshold), a new \SDRClass is trained according to the proposed Cross-Domain \ac{ai} methodology and deployed as an \emph{xApp} in the \ac{bs}. The newly trained classifier takes on the task of online traffic classification in the \ac{ran}. Subsequently, a new cycle of data collection and model training starts. The data gathered during this online classification cycle is appended to the \emph{RAN Training Dataset} in Table~\ref{tab:dtst} at the end of the experiment.

In Figure~\ref{fig:lifecycle}, the accuracy of our two classifiers is depicted over time. The blue line illustrates the accuracy of our \SDNClass, while the orange line represents the \SDRClass in each defined time period ($t(x)$). 
Notably, it is evident that the accuracy of the \SDRClass plateaus, a trend observable particularly in chunk from $t7$ to $t8$. Considering that this classifier is trained from feedback received from the \SDNClass, it is expected that its accuracy will never surpass the accuracy of the \SDNClass.

\vspace{-0.0cm}
\subsection{Classification Results} \label{sec:case_study:class_result}

To assess the accuracy of our \SDRClass, we use the model published in \cite{xavier2023ICC} as a \benchmark. This model was trained on the whole dataset, including the data collected in $t0$. Then we take $70\%$ of the entire dataset for training, and the remaining $30\%$ is used for testing. Furthermore, the labels of the datasets used for training the \SDRClass and \benchmark are not the same. The \benchmark was trained using real labels directly obtained from the traffic generator tools. On the other hand, the \SDRClass relies on feedback provided by the \SDNClass, which introduces noise due to mislabeling. However, it is important to notice that the \benchmark is not realistic, due to the fact that in a real network deployment, access to original traffic labels is not available and even if a dataset was collected, such an approach does not enable online model training over time. 

Table~\ref{tab:class_result} and Figure~\ref{fig:classConfMatrix} summarize the classification result of \benchmark, \SDRClass, and \SDNClass, which achieves the overall accuracy around $96\%$, $93\%$, and $98\%$ respectively. It is important to note that the \benchmark represents the maximum accuracy achievable by the \SDRClass. However, the \SDRClass has to deal with the mislabeling rate present in the \SDNClass. This introduces an error rate, evident in Figures \ref{fig:sdr_cm} and \ref{fig:sdn_cm}.


\vspace{-0.35cm}
\section{Conclusion and Future Work}
This paper has presented a Cross-Domain \ac{ai} training methodology for \ac{ml} traffic classification solutions for attack detection and mitigation, encompassing the \ac{ran} and the transport network domains. Specifically, the proposed methodology leverages the knowledge of an in-network \ac{ml} traffic classifier for \ac{sdn} switches in the transport network to support the training of an ML traffic classifier for the \ac{ran}, which can be deployed as an \emph{xApp} in the \ac{bs}. 

The methodology has been assessed in a realistic experimental setup deployed in OpenIreland testbed facilities. Results showed that the \SDRClass achieves lower accuracy than the \SDNClass. Despite this, the ability to identify the malicious flow near the source of the attack holds paramount importance. This proximity allows for the prevention of widespread network dissemination of attacks by enabling the timely interruption of the traffic, thus containing potential threats effectively. Likewise, the \benchmark presented higher accuracy than the \SDRClass. However, as previously mentioned, the \benchmark has unrealistic assumptions, like access to real traffic labels. This makes it static and does not allow for new traffic classes to be introduced in a fully automated way. The approach taken with the \SDRClass, on the other hand, is extremely useful for many network scenarios where datasets for proper training are not readily available. 

The decoupling between our \ac{ns} components (i.e., \ac{sdn} and \ac{ran} virtualization) empowers us to extend its applicability. Future work could explore transitioning from the knowledge encapsulated within the transport network and leverage insights from other systems such as firewalls, \ac{ids}), or statistical models. This flexibility highlights the adaptability and interconnectivity of our approach, enabling the use of diverse feedback mechanisms for comprehensive threat analysis.

\vspace{-0.1cm}

\section*{Acknowledgment}
{
Financial support from Science Foundation Ireland 17/CDA/4760, 18/RI/5721 and 13/RC/2077\_p2 is acknowledged. 
Financial support from Brazilian agencies: CNPq, CAPES, FAPESP/MCTI/CGI.br (PORVIR-5G 20/05182-3, and SAWI 20/05174-0), FAPES (94/2017, 281/2019, 515/2021, 284/2021, 1026/2022). CNPq fellows Dr. Martinello 306225/2020-4.
The work of Irene Vilà has been funded by European Union-NextGenerationEU, Spanish Ministry of Universities and the Plan for Recovery, Transformation, and Resilience, through the call for Margarita Salas Grants of the Universitat Politècnica de Catalunya (ref. 2022UPC-MSC- 94079).
}
\vspace{-0.1cm}

\bibliographystyle{IEEEtran}
\bibliography{main.bib}

\begin{thebibliography}{10}
\providecommand{\url}[1]{#1}
\csname url@samestyle\endcsname
\providecommand{\newblock}{\relax}
\providecommand{\bibinfo}[2]{#2}
\providecommand{\BIBentrySTDinterwordspacing}{\spaceskip=0pt\relax}
\providecommand{\BIBentryALTinterwordstretchfactor}{4}
\providecommand{\BIBentryALTinterwordspacing}{\spaceskip=\fontdimen2\font plus
\BIBentryALTinterwordstretchfactor\fontdimen3\font minus \fontdimen4\font\relax}
\providecommand{\BIBforeignlanguage}[2]{{%
\expandafter\ifx\csname l@#1\endcsname\relax
\typeout{** WARNING: IEEEtran.bst: No hyphenation pattern has been}%
\typeout{** loaded for the language `#1'. Using the pattern for}%
\typeout{** the default language instead.}%
\else
\language=\csname l@#1\endcsname
\fi
#2}}
\providecommand{\BIBdecl}{\relax}
\BIBdecl

\bibitem{chinese2023world}
C.~A. of~Cyberspace Studies xiaxueping@ cac.~gov. cn, ``World cybersecurity development,'' in \emph{World Internet Development Report 2021: Blue Book for World Internet Conference}.\hskip 1em plus 0.5em minus 0.4em\relax Springer, 2023, pp. 157--172.

\bibitem{popescu2022network}
A.~Popescu and A.~Westerhagen, ``Network softwarization: Developments and challenges,'' in \emph{2022 14th International Conference on Communications (COMM)}.\hskip 1em plus 0.5em minus 0.4em\relax IEEE, 2022, pp. 1--6.

\bibitem{kaur2021review}
S.~Kaur, K.~Kumar, and N.~Aggarwal, ``A review on p4-programmable data planes: Architecture, research efforts, and future directions,'' \emph{Computer Communications}, vol. 170, pp. 109--129, 2021.

\bibitem{qin2020learning}
Q.~Qin, K.~Poularakis, and L.~Tassiulas, ``A learning approach with programmable data plane towards iot security,'' in \emph{2020 IEEE 40th International Conference on Distributed Computing Systems (ICDCS)}.\hskip 1em plus 0.5em minus 0.4em\relax IEEE, 2020, pp. 410--420.

\bibitem{qin2022bringing}
------, ``Bringing intelligence at the network data plane for internet of things security,'' \emph{IoT for Defense and National Security}, pp. 259--283, 2022.

\bibitem{sha2023machine}
A.~Sha, S.~Madhan, S.~Neemkar, V.~B.~C. Varma, and L.~S. Nair, ``Machine learning integrated software defined networking architecture for congestion control,'' in \emph{2023 International Conference on Distributed Computing and Electrical Circuits and Electronics (ICDCECE)}.\hskip 1em plus 0.5em minus 0.4em\relax IEEE, 2023, pp. 1--5.

\bibitem{sacco2023hint}
A.~Sacco, A.~Angi, F.~Esposito, and G.~Marchetto, ``Hint: Supporting congestion control decisions with p4-driven in-band network telemetry,'' in \emph{2023 IEEE 24th International Conference on High Performance Switching and Routing (HPSR)}.\hskip 1em plus 0.5em minus 0.4em\relax IEEE, 2023, pp. 83--88.

\bibitem{liu2023load}
W.-X. Liu, J.~Cai, Y.-H. Zhu, J.-M. Luo, and J.~Li, ``Load balancing inside programmable data planes based on network modeling prediction using a gnn with network behaviors,'' \emph{Computer Networks}, vol. 227, p. 109695, 2023.

\bibitem{alhilali2023artificial}
A.~H. Alhilali and A.~Montazerolghaem, ``Artificial intelligence based load balancing in sdn: A comprehensive survey,'' \emph{Internet of Things}, p. 100814, 2023.

\bibitem{wu2023p4sqa}
Q.~Wu, Q.~Liu, Z.~Jia, N.~Xin, and T.~Chen, ``P4sqa: A p4 switch-based qos assurance mechanism for sdn,'' \emph{IEEE Transactions on Network and Service Management}, 2023.

\bibitem{keshari2021systematic}
S.~K. Keshari, V.~Kansal, and S.~Kumar, ``A systematic review of quality of services (qos) in software defined networking (sdn),'' \emph{Wireless Personal Communications}, vol. 116, pp. 2593--2614, 2021.

\bibitem{ali2022qos}
J.~Ali, M.~Adnan, T.~R. Gadekallu, R.~H. Jhaveri, and B.-H. Roh, ``A qos-aware software defined mobility architecture for named data networking,'' in \emph{2022 IEEE Globecom Workshops (GC Wkshps)}.\hskip 1em plus 0.5em minus 0.4em\relax IEEE, 2022, pp. 444--449.

\bibitem{zhu2022machine}
X.~Zhu and Y.~Zhang, ``Machine-learning-assisted traffic classification of user activities at programmable data plane,'' in \emph{2022 23rd Asia-Pacific Network Operations and Management Symposium (APNOMS)}.\hskip 1em plus 0.5em minus 0.4em\relax IEEE, 2022, pp. 01--04.

\bibitem{paramasivam2023cor}
S.~Paramasivam and R.~L. Velusamy, ``Cor-entc: correlation with ensembled approach for network traffic classification using sdn technology for future networks,'' \emph{The Journal of Supercomputing}, vol.~79, no.~8, pp. 8513--8537, 2023.

\bibitem{xavier2022map4}
B.~M. Xavier, R.~S. Guimar{\~a}es, G.~Comarela, and M.~Martinello, ``Map4: A pragmatic framework for in-network machine learning traffic classification,'' \emph{IEEE Transactions on Network and Service Management}, vol.~19, no.~4, pp. 4176--4188, 2022.

\bibitem{musumeci2022machine}
F.~Musumeci, A.~C. Fidanci, F.~Paolucci, F.~Cugini, and M.~Tornatore, ``Machine-learning-enabled ddos attacks detection in p4 programmable networks,'' \emph{Journal of Network and Systems Management}, vol.~30, pp. 1--27, 2022.

\bibitem{zang2023towards}
M.~Zang, C.~Zheng, L.~Dittmann, and N.~Zilberman, ``Towards continuous threat defense: in-network traffic analysis for iot gateways,'' \emph{IEEE Internet of Things Journal}, 2023.

\bibitem{sacco2023p4fl}
A.~Sacco, A.~Angi, G.~Marchetto, and F.~Esposito, ``P4fl: An architecture for federating learning with in-network processing,'' \emph{IEEE Access}, 2023.

\bibitem{bonati2021intelligence}
L.~Bonati, S.~D'Oro, M.~Polese, S.~Basagni, and T.~Melodia, ``Intelligence and learning in {O-RAN} for data-driven nextg cellular networks,'' \emph{IEEE Communications Magazine}, vol.~59, no.~10, pp. 21--27, 2021.

\bibitem{ferrag2022edge}
M.~A. Ferrag, O.~Friha, D.~Hamouda, L.~Maglaras, and H.~Janicke, ``Edge-iiotset: A new comprehensive realistic cyber security dataset of iot and iiot applications for centralized and federated learning,'' \emph{IEEE Access}, vol.~10, pp. 40\,281--40\,306, 2022.

\bibitem{vamsi2022detailed}
K.~Vamsi~Krishna, K.~Swathi, P.~Rama Koteswara~Rao, and B.~Basaveswara~Rao, ``A detailed analysis of the cidds-001 and cicids-2017 datasets,'' in \emph{Pervasive Computing and Social Networking: Proceedings of ICPCSN 2021}.\hskip 1em plus 0.5em minus 0.4em\relax Springer, 2022, pp. 619--638.

\bibitem{xavier2023ICC}
B.~M. Xavier, M.~Dzaferagic, D.~Collins, G.~Comarela, M.~Martinello, and M.~Ruffini, ``{Machine Learning-based Early Attack Detection Using Open RAN Intelligent Controller},'' in \emph{ICC 2023-IEEE International Conference on Communications}.\hskip 1em plus 0.5em minus 0.4em\relax IEEE, 2023.

\bibitem{pinheiro20235geco}
J.~F.~N. Pinheiro, C.-Y. Chang, T.~Collins, E.~Smekens, R.~Berozashvili, A.~Shahid, D.~De~Vleeschauwer, P.~Soto, I.~Moerman, J.~Marquez-Barja \emph{et~al.}, ``5geco: A cross-domain intelligent neutral host architecture for 5g and beyond,'' in \emph{IEEE INFOCOM 2023-IEEE Conference on Computer Communications Workshops (INFOCOM WKSHPS)}.\hskip 1em plus 0.5em minus 0.4em\relax IEEE, 2023, pp. 1--6.

\bibitem{10195597}
R.~Ferreira, J.~Fonseca, J.~Silva, M.~Tendulkar, P.~Duarte, M.~Araújo, R.~Barbosa, B.~Mendes, and A.~Goes, ``Demo: Enhancing network performance based on 5g network function and slice load analysis,'' in \emph{2023 IEEE 24th International Symposium on a World of Wireless, Mobile and Multimedia Networks (WoWMoM)}, 2023, pp. 340--342.

\bibitem{alliance2021ran_WG2}
O.~Alliance, ``{O-RAN Working Group 2. AI/ML Workflow Description and Requirements (O-RAN. WG2. AIML-v01. 02)},'' \emph{Tech. Rep}, 2021.

\bibitem{nguyen2020efficient}
T.~G. Nguyen, T.~V. Phan, D.~T. Hoang, T.~N. Nguyen, and C.~So-In, ``Efficient sdn-based traffic monitoring in iot networks with double deep q-network,'' in \emph{International conference on computational data and social networks}.\hskip 1em plus 0.5em minus 0.4em\relax Springer, 2020, pp. 26--38.

\bibitem{zhou2020flow}
Y.~Zhou, C.~Sun, H.~H. Liu, R.~Miao, S.~Bai, B.~Li, Z.~Zheng, L.~Zhu, Z.~Shen, Y.~Xi \emph{et~al.}, ``Flow event telemetry on programmable data plane,'' in \emph{Proceedings of the Annual conference of the ACM Special Interest Group on Data Communication on the applications, technologies, architectures, and protocols for computer communication}, 2020, pp. 76--89.

\bibitem{barradasflowlens}
D.~Barradas, N.~Santos, L.~Rodrigues, S.~Signorello, F.~M. Ramos, and A.~Madeira, ``Flowlens: Enabling efficient flow classification for ml-based network security applications,'' in \emph{Proceedings of the Network and Distributed Systems Security (NDSS) Symposium}, 2021.

\bibitem{alliance2021ran_WG3_RC}
O.~Alliance, ``{O-RAN Working Group 3: Near-Real-time RAN Intelligent Controller E2 Service Model, RAN Control 1.0 (ORAN-WG3. E2SM-RC-v01.00)},'' \emph{Tech. Spec}, 2021.

\bibitem{sharafaldin2018toward}
I.~Sharafaldin \emph{et~al.}, ``Toward generating a new intrusion detection dataset and intrusion traffic characterization.'' in \emph{ICISSP}, 2018, p. 108.

\end{thebibliography}

\end{document}